# Convergence of Symbiotic Communications and Blockchain for Sustainable and Trustworthy 6G Wireless Networks

Haoxiang Luo, Gang Sun, *Senior Member, IEEE*, Cheng Chi, Hongfang Yu, *Senior Member, IEEE*, and Mohsen Guizani, *Fellow, IEEE*

*Abstract*—Symbiotic communication (SC) is known as a new wireless communication paradigm, similar to the natural ecosystem population, and can enable multiple communication systems to cooperate and mutualize through service exchange and resource sharing. As a result, SC is seen as an important potential technology for future sixth-generation (6G) communications, solving the problem of lack of spectrum resources and energy inefficiency. Symbiotic relationships among communication systems can complement radio resources in 6G. However, the absence of established trust relationships among diverse communication systems presents a formidable hurdle in ensuring efficient and trusted resource and service exchange within SC frameworks. To better realize trusted SC services in 6G, in this paper, we propose a solution that converges SC and blockchain, called a symbiotic blockchain network (SBN). Specifically, we first use cognitive backscatter communication to transform blockchain consensus, that is, the symbiotic blockchain consensus (SBC), so that it can be better suited for the wireless network. Then, for SBC, we propose a highly energy-efficient sharding scheme to meet the extremely low power consumption requirements in 6G. Finally, such a blockchain scheme guarantees trusted transactions of communication services in SC. Through ablation experiments, our proposed SBN demonstrates significant efficacy in mitigating energy consumption and reducing processing latency in adversarial networks, which is expected to achieve a sustainable and trusted 6G wireless network.

## I. Introduction

As the commercial deployment of fifth-generation (5G) communication technologies gains momentum, the research and development endeavors in anticipation of sixth-generation (6G) communication systems have captured the attention of numerous scholars. Citing a report by Ericsson [1], projections indicate that by 2027, global mobile network traffic will surge to approximately 370 EB per month, surpassing 30.2 billion Internet of Things (IoT) connections. The forthcoming deluge of data transmission and proliferation of end-user terminals pose significant challenges for 6G network architecture and design. Primarily, the quest for additional spectrum resources to meet the escalating demands of wireless traffic and terminal connectivity poses a critical hurdle, exacerbated by the near exhaustion of optimal low and medium frequency bands reserved for wireless communication. This spectrum scarcity scenario poses a substantial bottleneck impeding the prospective development of 6G networks. Moreover, the information and communication sector already contributes 2% to 4% of global carbon emissions [2], underlining the environmental concerns associated with the industry's energy consumption. The anticipated surge in wireless traffic volumes and terminal connections further exacerbates the burgeoning energy consumption concerns plaguing 6G network deployment, directly contradicting the envisioned paradigm of ultra-low-power communication protocols inherent to the 6G vision [3].

In response to the aforementioned challenges, symbiotic communication (SC) has emerged as a prospective remedy [2], aiming to orchestrate collaborative resource sharing among diverse radios to enhance resource utilization efficiency. Conceptually akin to symbiotic relationships observed in natural ecosystems, SC mirrors synergistic alliances within communication systems through mutualistic resource sharing and service exchange. Unlike conventional network paradigms characterized by pre-allocated communication resources and restricted service provider interactions, SC envisions the entirety of the 6G network as a dynamic radio ecosystem, fostering a collaborative evolution among all symbiotic radio devices (SRDs) through continuous service and resource transactions. Consequently, resource constraints within the network can be mitigated as disparate resource bottlenecks are alleviated through the exchange of diverse resources and services. For instance, SC can incentivize collaboration between satellites and ground terminals with varied resources to synergistically enhance the network's overall performance [4].

Nevertheless, the integration of SC within the infrastructure of 6G networks encounters a critical obstacle about the establishment of trust relationships among individual SRDs [5]. The absence of trust among SRDs poses a formidable challenge, impeding their ability to make informed decisions regarding service and resource transactions, potentially resulting in unfair and opaque transactions. Such shortcomings could impact secure and smooth operations of 6G networks. The inherent challenges arise from factors such as extensive transmission distances, the dynamic network environment, and the intricate electromagnetic interference landscape, leading to

H. Luo, G. Sun (corresponding author), and H. Yu, are with the University of Electronic Science and Technology of China, China (e-mail: lhx991115 @163.com; {gangsun, yuhf}@ uestc.edu.cn).
C. Chi is with the China Academy of Information and Communications Technology, China (e-mail: chicheng@caict.ac.cn).
M. Guizani is with the Mohamed Bin Zayed University of Artificial Intelligence (MBZUAI), UAE (e-mail: mguizani@ieee.org).

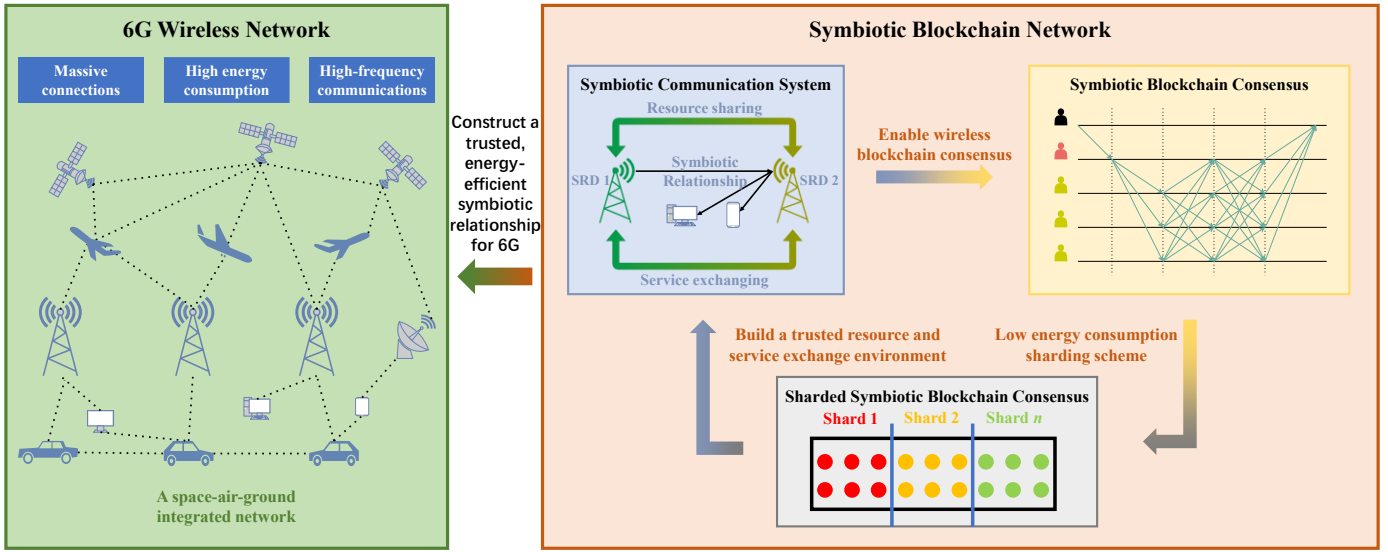

**Fig. 1.** Concrete steps for our sustainable and trustworthy 6G wireless network.

prevalent instances of unreliable information exchanges even in the absence of external attacks. Moreover, the presence of Byzantine or faulty SRDs further compounds these issues. In the absence of an effective consensus mechanism, the realization of trusted transactions becomes extremely difficult, and there is an urgent need to tailor a secure transaction solution for SC deployed in 6G networks.

Fortuitously, blockchain presents a viable solution to the challenges above. As a decentralized and immutable distributed ledger, blockchain leverages smart contracts and cryptographic protocols to facilitate secure, reliable, and automated execution of service and resource transactions [6]. It has been widely deployed in various wireless networks [7-8]. Notably, the consensus mechanism inherent in blockchain empowers network entities to achieve consensus autonomously, circumventing the need for intervention by third-party trust parties. As a result, implementing blockchain within 6G networks fosters the establishment of robust and trustworthy symbiotic relationships. However, there are still few efforts to integrate SC and blockchain, i.e. [5] and [9]. In particular, for 6G networks enabled by SC, the existing blockchain consensus design generally has the following issues.

**Not applicable to the SC:** The convergence of SC and blockchain not only signifies the capability of blockchain to facilitate trusted service and resource transactions within SC networks, but also underscores the potential for SC to optimize blockchain consensus mechanisms for enhanced applicability within 6G wireless environments.

**Without consideration for 6G features:** Beyond the essential requirement for secure transactions in SC, blockchain consensus mechanisms must align with key tenets of 6G networks, including ultra-low energy communication and massive connectivity. These considerations directly address the challenges related to consensus energy consumption and scalability.

These motivations are illustrated in Fig. 1, which shows the concrete steps to build our sustainable and trustworthy 6G wireless networks. This figure uses the space-air-ground integrated network as an example of a future 6G wireless network scenario. Based on this, our novel proposition of a Symbiotic Blockchain Network (SBN) that interweaves SC and blockchain elements propels us toward realizing a sustainable and trustworthy 6G wireless network. In particular, the notable contributions of this study are summarized as follows.

- We use cognitive backscatter communication in SC to construct mutualistic transmission relationships for nodes in a wireless blockchain network (WBN) to design a symbiotic blockchain consensus (SBC) with low energy consumption and high reliability.
- We design a low-energy sharding scheme to amplify the energy efficiency and scalability of the proposed SBC, thereby optimizing its utility within 6G wireless networks.
- We implement the sharded SBC in an SC-enabled 6G wireless network to construct the SBN, supporting efficient, sustainable, and trustworthy transactions of resources and services within the SC.

## II. SYMBIOTIC COMMUNICATION

In this section, we briefly introduce SC, divided into obligate and facultative symbiosis. Then, we present the framework of cognitive backscatter communication, the most typical example of SC.

### A. Symbiotic Relationship

Organisms in nature consume a variety of resources (including food, light, etc.) to accomplish specific tasks such as protection and feeding. Similarly, SRDs in SC consume communication resources such as spectrum, time, and energy to relay, calculate, and transmit [2].

Following the mutually beneficial and win-win population relationship in nature, many researchers also hope that different communication systems can establish the exchange of communication tasks such as relay and transmission, as well as the sharing of spectrum and energy resources, to achieve the symbiotic relationship between communication systems, namely the SC.

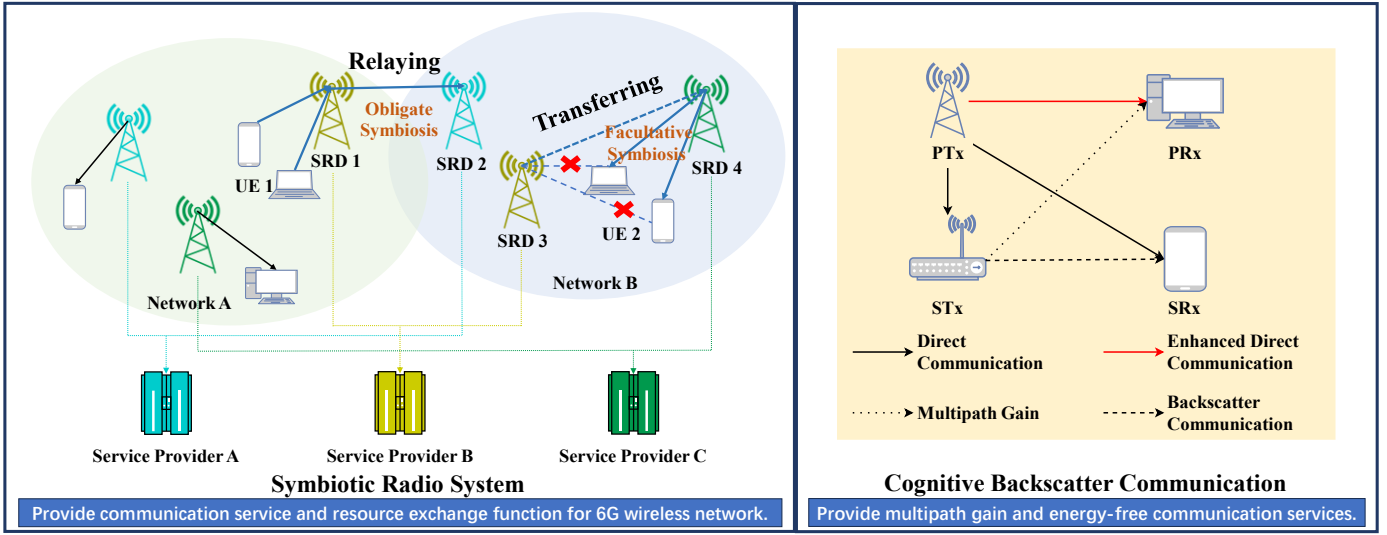

**Fig. 2.** Symbiotic communication.

In such a communication paradigm, all SRDs are expected to achieve communication performance gains through the exchange of resources and services. Specifically, symbiotic relationships can be divided into obligate and facultative relationships [2].

The obligate relationship pertains to the scenario where an SRD heavily depends on the collaborative efforts of other SRDs to furnish communication services to user equipment (UE), as it cannot independently achieve its communication objectives. For instance, cognitive backscatter communication is an example of such interdependence, which we shall delve into further in subsequent discussions. Illustrated in the left side of Fig. 2, the incapacity of SRD 2 to deliver network access services to UE 1 without the intermediary support from SRD 1 exemplifies this inherent reliance. This relationship is very similar to the plant-bee relationship, where plants furnish bees with essential pollen as sustenance, and reciprocally, bees aid in pollinating flowers. Neither entity can thrive autonomously in the natural habitat, thereby qualifying as obligate symbiosis.

In addition, facultative symbiosis describes that each SRD can perform communication tasks as an independent server, but together they can provide higher-quality communication services to UEs. As depicted on the left side of Fig. 2, both SRD 3 and SRD 4 can independently provide network access services for the UE 2. When one SRD is unavailable, another SRD can serve the UE 2. And these SRDs can provide better communication services for UE through resource-sharing. The relationship is similar to that between sharks and remoras. Remoras receive additional nourishment by cleaning food remnants from the shark's teeth and removing parasites from its skin. This mutually beneficial arrangement, known as facultative symbiosis, allows the two species to coexist harmoniously despite being able to survive independently of each other.

*B. Cognitive Backscatter Communication*

Cognitive backscatter communication is the most widely used communication paradigm in SC [10]. In WBNs, SC can achieve efficient spectrum utilization, energy saving, and high communication reliability [11].

Specifically, cognitive backscatter communication includes a primary communication system and a secondary communication system. The primary communication system incorporates the primary transmitter (PTx) and the primary receiver (PRx), while the secondary communication system is composed of the secondary transmitter (STx) and the secondary receiver (SRx). With proper deployment, STx can provide multipath gain to PRx, thereby enhancing the communication reliability of the primary communication system. Notably, SRx can obtain information from STx utilizing the radio frequency (RF) signal emitted by PTx, without additional energy consumption. This system presents a promising avenue for mitigating communication reliability and energy consumption challenges in WBNs, as illustrated on the right side of Fig. 2, with design details expounded in [12].

### III. SYMBIOTIC BLOCKCHAIN NETWORK

Considering the challenges of blockchain and SC, we integrate and propose the SBN framework, as shown in Fig. 3. This section is divided into three parts to describe it, the symbiotic services provided for SC, the role of blockchain, and low-energy-consumption sharding.

*A. Symbiotic Services in SBN*

Within the conceptual framework of the elucidated SBN, the network entities encompass UEs and SRDs, with the latter category encompassing an array of components such as satellites, aircraft or drones, base stations (BSs), and vehicles, among others. The symbiotic services primarily unfold among the SRDs, facilitating reciprocal service or resource exchanges among neighboring entities, and enabling the provisioning of network access services to UEs encompassed within their respective coverage areas.

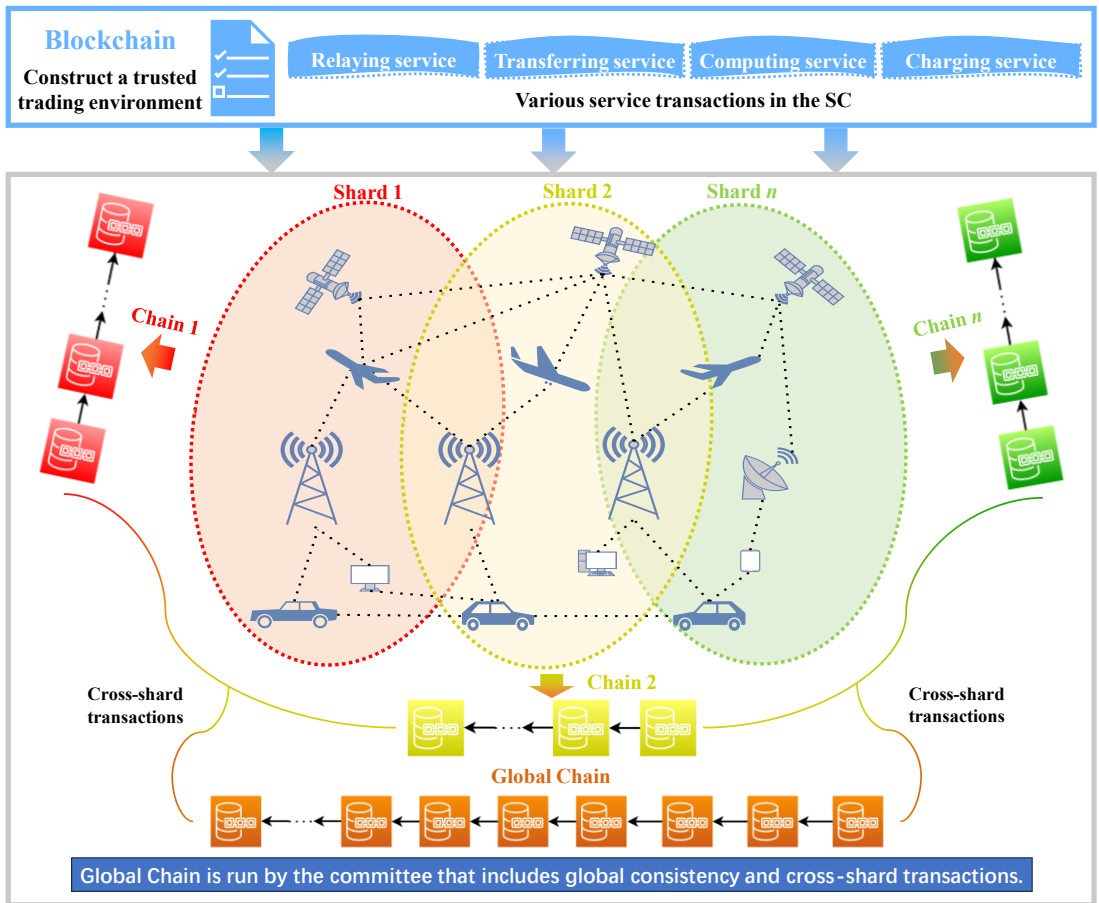

**Fig. 3.** Framework of the symbiotic blockchain network.

Inspired by [9], SBN facilitates the exchange of four distinct services, encompassing the collaborative sharing of three essential resources: spectrum, computing, and energy, as delineated below:

**Relaying service:** In instances where the UE encounters difficulty directly connecting to the network serviced by SRD B, necessitating access, SRD A leverages its spectrum resources to relay radio signals for the UE. This collaborative relationship engenders an obligate symbiosis between SRD A and SRD B, since UE cannot access the network provided by SRD B without SRD A.

**Transferring service:** In the event that SRD A is unable to furnish the necessary network services for a UE, the responsibility of network access is seamlessly transitioned to SRD B. Consequently, the network access point shifts from SRD A to SRD B for the UE. This service transfer can give rise to a facultative relationship between the SRDs involved, as both can jointly provide network access services for the UE.

**Computing service:** In scenarios where the computing capabilities of SRD A are inadequate, SRD B can facilitate supplementary computing support through task offloading. This collaborative relationship between SRD A and SRD B may exhibit characteristics of either obligate or facultative symbiosis, depending on the specific context and level of interdependence. Specifically, when SRD A faces computational overload and SRD B collaboratively accomplishes the computation task, this symbiotic relationship is considered facultative. Conversely, if SRD A lacks sufficient computing power, resulting in SRD B assuming sole responsibility for task completion, it exemplifies an obligate symbiosis.

**Charging service:** In cases where SRD A experiences power capacity limitations, SRD B possesses the capability to augment its power resources, facilitating the successful completion of communication tasks. This collaboration can be accomplished through power replacement, integrated data and energy transfer, etc. Through the power support services, an obligate or facultative relationship can be fostered between the two SRDs. The distinction between the two forms of symbiosis lies in the fact that the former necessitates SRD A to rely on the power supply from SRD B to fulfill the communication task, whereas the latter allows SRD A to access additional electric energy for expeditious and enhanced task completion.

It should be noted that SRDs A and B mentioned here have no specific correspondence in figures, and only serve as indications here.

*B. Blockchain Consensus in SBN*

The primary function of blockchain within the SBN is to establish a secure transaction framework for the symbiotic services featured in the 6G wireless network, ensuring trust and integrity. Additionally, blockchain must facilitate an effective verification mechanism for these transactions. The detailed implementation of this dual role is outlined below:

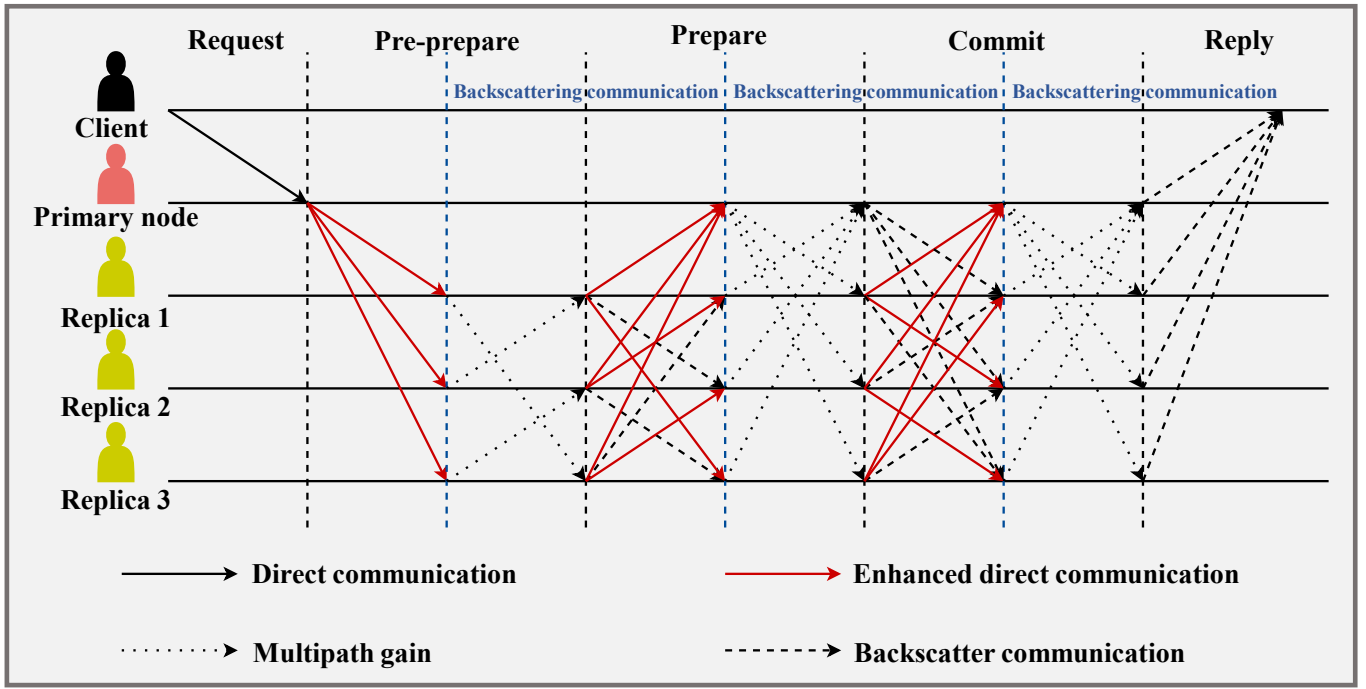

**Fig. 4.** Consensus process of S-PBFT.

**Trusted trading environment:** In 6G wireless networks, SRDs may originate from distinct network operators, lacking inherent trust amongst them. Thus, establishing a reliable environment for service transactions between SRDs becomes imperative. The blockchain consensus mechanism serves as an evaluative tool pivotal in network decision-making, enabling all SRDs to independently assess transaction legitimacy based on factors such as SRD balance of account, transaction amount, and timing, devoid of third-party intervention. Upon unanimous consistency among SRDs, transactions are securely inscribed onto the blockchain as hash values, ensuring reliability, security, and traceability.

**Efficient transaction processing and verification:** The interconnection among SDRs facilitates the provision of ubiquitous symbiotic services within the SBN. Nevertheless, these transactions typically encompass time-sensitive characteristics, necessitating expedient processing. Consequently, the blockchain implementation within the SBN must have high scalability and rapid transaction verification capability. We adopt sharding in the SBN to meet this need to enable parallel processing and verification of all transactions. Each shard maintains its sub-chains separately. The data structure of transactions includes the identifiers of the transaction demander and the provider to determine which shard the transaction comes from, thereby enhancing traceability efficiency. The sharding details will be revealed in Part C of this section.

As indicated above, the utilization of blockchain to establish a dependable transaction environment for symbiotic services, alongside the sharding-based transaction processing and verification mechanisms, are contingent on robust blockchain consensus. In WBNs, path loss, channel fading, and varying environmental conditions often affect the consensus, resulting in a low consensus success rate, especially in 6G [13]. Moreover, in wireless scenarios, SRDs are difficult to obtain a timely power supply. Therefore, we try to integrate cognitive backscatter communication into WBNs to introduce a novel SBC. This innovative solution aims to alleviate the unreliable communication and high energy consumption issues in wireless consensus. By leveraging Practical Byzantine Fault Tolerance (PBFT) as a case study, we demonstrate that reasonable setting of consensus nodes as primary or secondary communication systems can enhance the consensus success rate by 54.1% and diminish consensus energy consumption by 9.2% [11]. Fig.4 shows the PBFT consensus after cognitive backscatter communication modification, which is called S-PBFT. The dashed lines symbolize the backscatter communication transmitted by STx utilizing the RF signals from PTx, wherein a portion serves as a multipath gain for the primary communication system, while another part is dedicated to transmitting consensus information to SRx without energy expenditure. And the design details can be found in [11].

*C. Low-energy-consumption Sharding for SBN*

Although SBC can mitigate high energy consumption through symbiosis, the blockchain tends to be an energy-intensive component. In addition, the blockchain consensus also faces challenges related to low scalability, rendering it less suitable for meeting the extensive networking demands characteristic of 6G [3], and inadequate to fulfill efficient transaction processing and verification. Therefore, we plan to further design a low-energy sharding scheme for SBC to completely solve the above issues.

In SBNs, the SRD except a BS is mostly mobile. Therefore, we take the BS as the target and design the sharding scheme. Drawing inspiration from [13-14], we construct an energy consumption model $E$ for S-PBFT based on the parameters of BS number $Z$, shard quantity $n$, BS count per shard $m$, and the path loss exponent $\alpha$. Other SRDs are added to the shard where the BS resides. Subsequently, by computing the second derivative of this model, we discern a consistent positivity in the second derivative, which indicates that the first derivative of energy consumption increases monotonically within the range of the above parameters. Consequently, we set the first derivative to zero to ascertain the optimal sharding configuration within S-PBFT, i.e. determine the values of $n$, and $m$, thereby minimizing consensus energy consumption. These steps details are shown in Fig. 5.

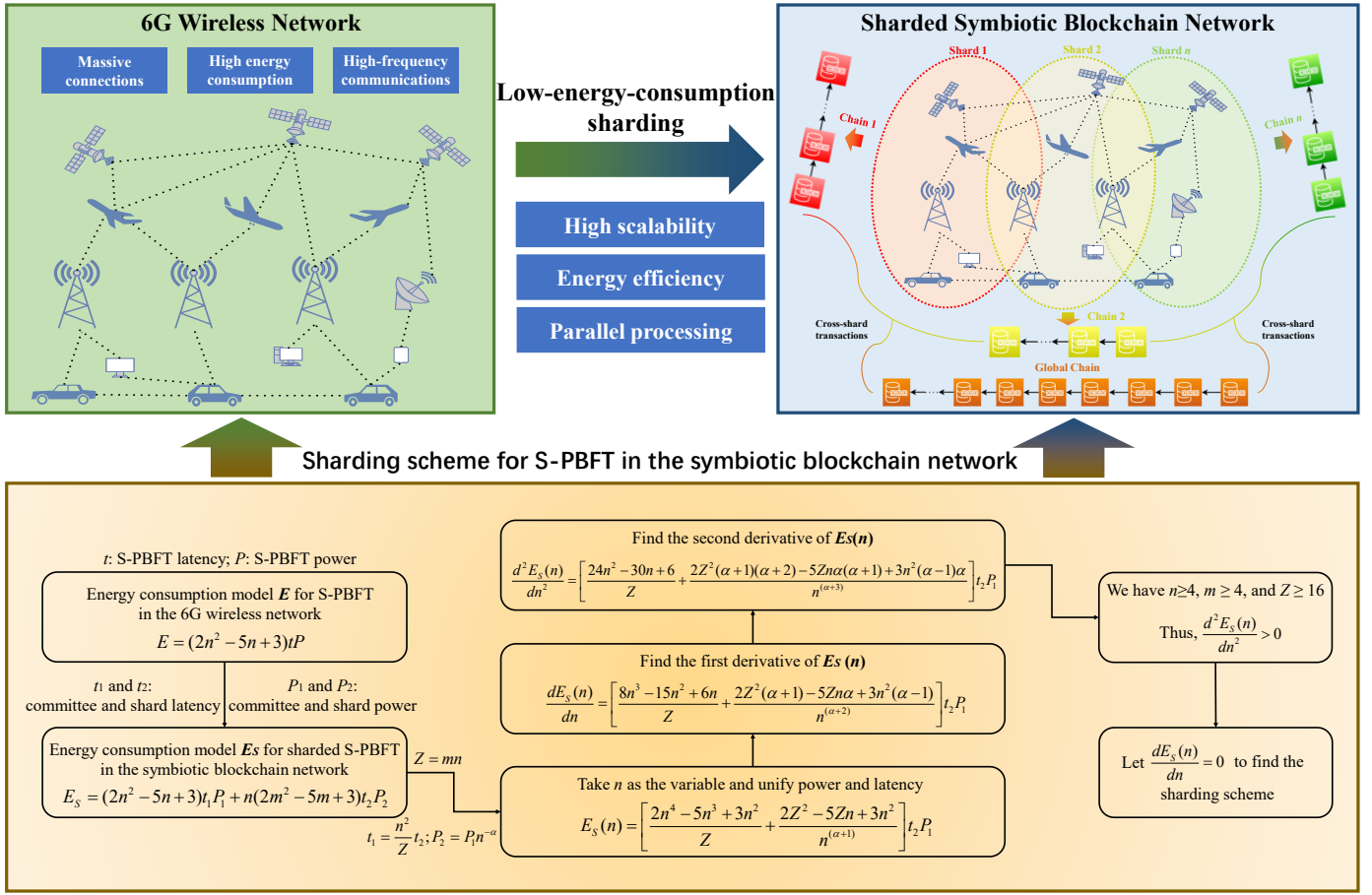

**Fig. 5.** Sharding scheme for S-PBFT in the symbiotic blockchain network.

Since BS remains stationary within our proposed sharding framework, a BS is selected as the leader within each shard based on criteria such as reputation value and activity level. The designated lead BS assumes the role of initiating consensus among other SRDs within the same shard. Subsequently, the lead BS endeavors to establish final global consistency by forming a committee in collaboration with other lead BSs across different shards. If a cross-shard symbiotic service occurs, it will be coordinated by this committee. Global consistency and cross-chain transactions will be packaged and linked to the committee responsible chain, called Global Chain.

## IV. CASE STUDY

### A. Simulation Settings

In this section, we conduct ablation tests to assess the communication functionalities of the SBN, illustrating the positive impact of SBC, low-energy consumption sharding, and symbiotic services on enhancing the performance of 6G wireless networks. Additionally, these simulations are evaluated across three distinct network scenarios:

**NA:** there is no SRD as the attacker.

**FA:** 10% of SRDs are fault attackers who do not participate in the symbiotic service exchange and consensus process.

**BA:** 10% of SRDs are Byzantine attackers who initiate forged transactions in the symbiotic service exchange and consensus process.

Then, we examine a scenario of 6G wireless networks comprising 50 BSs, 20 UAVs, 20 ground mobile SRDs, and 10 satellites. Each SRD possesses a communication range of 200 m, while the ground mobile SRDs exhibit a mobile radius of 150 m. The flight and communication parameters for UAVs and satellites are determined by referencing a previous study [15]. Furthermore, the communication bandwidth between each SRD is set at 20 MHz, with a path loss exponent $\alpha=3$. The network configuration involves 100 SRDs catering to network access services for 150 UEs. Each UE generates communication requests at a rate of 20 Mbps, with a latency requirement of 50 ms.

Fig. 6 illustrates the energy consumption and latency metrics observed within the SBN when processing UE communication requests. The results are derived from an average of 200 simulations, which are conducted using MATLAB R2021a on a PC outfitted with an Intel I7-1260P processor boasting a clock frequency of 2.1 GHz and 16 GB

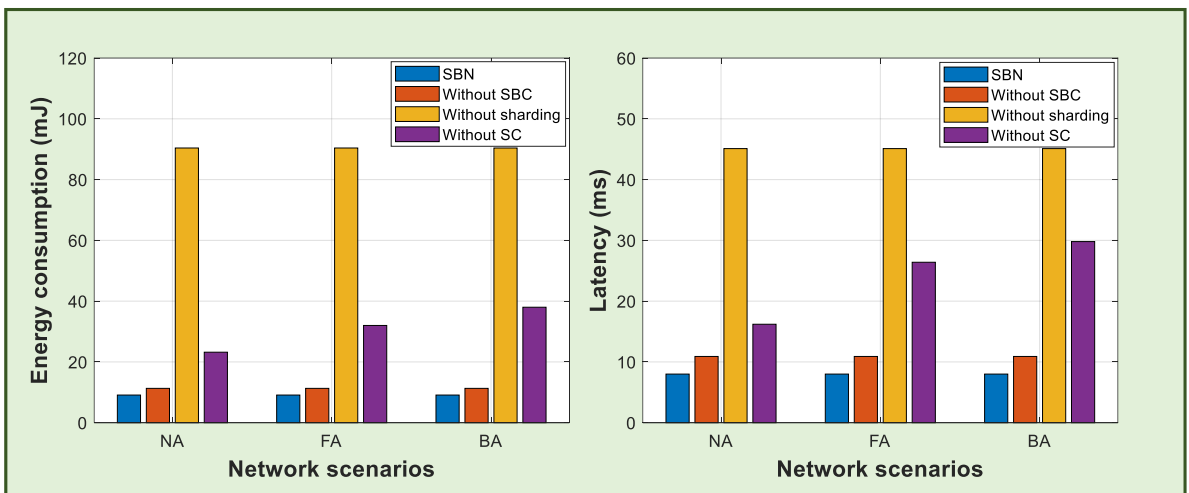

**Fig. 6.** Simulation results of energy consumption and latency.

RAM. Within the experimental design, "SBN" denotes our proposed trusted and energy-efficient communication service framework tailored for the 6G wireless network. "Without SBC" signifies the conventional PBFT consensus framework lacking integration with SC. "Without sharding" indicates the absence of a sharding scheme in the blockchain configuration. Lastly, "Without SC" characterizes a network setting where radio devices do not engage in symbiotic relationships for service and resource exchange.

*B. Energy Consumption*

The ablation test depicted on the left side of Fig. 6 focuses on assessing the energy consumption within the SBN. The results demonstrate a notable increase in energy consumption in scenarios where neither SBC nor sharding, nor SC, are implemented, underscoring the positive impact of each module within SBN on energy efficiency. Specifically, sharding emerges as the most influential factor in reducing energy consumption, because it limits the communication negotiation of consensus to each shard, reducing the communication overhead and consensus latency of the SBN, and the transmit power of the SRD. Moreover, the inability of radio devices in SBN to establish symbiotic relationships results in heightened network energy consumption, emphasizing the energy-saving potential of symbiotic collaborations and resource complementarity. Meanwhile, the transformed consensus SBC offers energy-saving advantages, as backscatter communication replaces active communication which typically consumes more energy.

Furthermore, the energy consumption may display variability across different network scenarios. Notably, the SBN exhibits resilience against both types of attacks, showcasing its robust fault tolerance and ability to effectively counter Byzantine adversaries, thereby ensuring heightened security and fostering a trustworthy communication environment within 6G wireless networks. This reason comes from the fault-tolerant space provided by blockchain consensus, allowing some nodes in the network to fail or even be Byzantine. Additionally, the absence of sharding and SBC does not impact energy consumption levels, as these components primarily influence the consensus process and remain unaffected by attackers. Conversely, in the absence of SC, the energy consumption of SBN escalates under both attack scenarios, highlighting the impact of attackers on symbiotic services within the network.

*C. Latency*

The ablation test presented on the right side of Fig. 6 pertains to the latency metric within the SBN. Similar to energy consumption, the latency performance of SBN remains unaffected by varying network scenarios or types of attacks, underscoring its proficiency in efficiently handling communication tasks, even in the presence of adversarial elements. Notably, the absence of sharding results in a significant surge in latency, which is consistent with our hypothesis that sharding can reduce latency by limiting communication negotiation of consensus to each shard and processing transactions in parallel. This accentuates the pivotal role of sharding in enhancing the scalability of blockchain networks. Moreover, the integration of SBC and SC establishes a symbiotic relationship aimed at enabling blockchain consensus and communication services, consequently expediting the consensus process and service transactions. The above results show that the SBC, sharding, and symbiotic services designed in SBN have optimized effect on latency.

Additionally, latency experiences alteration solely when SC is absent, in the presence of attackers, exhibiting a pattern analogous to that observed in energy consumption modifications.

V. FUTURE DIRECTIONS

In this section, we outline three main future directions for implementing SBN.

*A. Transaction Models for Symbiotic Services*

While this study introduces a blockchain-enabled trusted transaction framework for symbiotic services, fostering symbiotic relationships among radio devices to enhance communication efficiency in 6G wireless networks, there exists an existing gap in the specific transaction modeling within this domain. To address this knowledge gap, researchers can begin by delving into fundamental theories such as game theory and queuing theory to establish a transaction model tailored for symbiotic services. This endeavor aims to refine the service efficiency within the SBN through model optimization.

*B. Abundant Symbiotic Relationships*

In our SBN, our analysis has predominantly focused on four core symbiotic services: relaying, transferring, computing, and charging. Nevertheless, these services may not comprehensively encapsulate all potential symbiotic relationships within the network. Furthermore, while mutual benefits can optimize outcomes for each radio device and the network, the potential role of competition in specific network scenarios warrants further investigation. In light of diverse network contexts and requirements, there arises a necessity to cultivate more comprehensive symbiotic relationships.

*C. Appropriate Blockchain Consensus Mechanisms*

In the SBN of this paper, cognitive backscatter communication is harnessed to revolutionize SBC, showcasing marked improvements in energy efficiency and consensus success rates compared to conventional wireless blockchain consensus mechanisms. Nonetheless, a suitable consensus mechanism tailored specifically for symbiotic service exchange remains unexplored, emphasizing the critical need for researchers to develop a consensus mechanism aligned with the unique requirements of 6G wireless networks and focused on symbiotic relationships.

## V. CONCLUSIONS

This study presents a novel approach for integrating blockchain and symbiotic communication in 6G wireless networks, namely SBN, thereby establishing a foundation for fostering trust and sustainability within the network infrastructure. Initially, we leverage cognitive backscatter communication to revolutionize wireless blockchain consensus, achieving a heightened consensus success rate coupled with reduced energy consumption. Subsequently, a low-energy sharding scheme is meticulously crafted for this consensus framework. This scheme is integrated into the symbiotic radio system to establish a trustworthy symbiotic relationship amongst radio devices within the 6G wireless network, accelerating the communication process through service and resource exchange. Through ablation experiments, it is verified that SBN can meet the above design requirements and effectively resist various attacks.

Overall, our endeavor aims to establish a groundwork for synergistic collaboration between blockchain and symbiotic communication, offering a promising solution towards building a sustainable and trustworthy 6G wireless network.